\documentclass[%
 reprint,
superscriptaddress,
 amsmath,amssymb,
 aps,
prx,longbibliography,
]{revtex4-2}
\usepackage{graphicx}% Include figure files
\usepackage{dcolumn}% Align table columns on decimal point
\usepackage{bm}% bold math
\usepackage{multirow}
\usepackage{float}
\newcommand\MyHead[2]{%
  \multicolumn{1}{l}{\parbox{#1}{\centering #2}}
}

\AtBeginDocument{%
	\newwrite\bibnotes
	\def\bibnotesext{SnSe.bib}
	\immediate\openout\bibnotes=\jobname\bibnotesext
	\immediate\write\bibnotes{@CONTROL{REVTEX42Control}}
	\immediate\write\bibnotes{@CONTROL{%
			apsrev42Control,author="08",editor="1",pages="1",title="0",year="1"}}
	\if@filesw
	\immediate\write\@auxout{\string\citation{apsrev42Control}}%
	\fi
}%
\begin{document}
\title{Observation of a Novel Lattice Instability in Ultrafast Photoexcited SnSe}% Force line breaks with \\
\author{Yijing Huang}
\affiliation{Stanford Institute for Materials and Energy Sciences, SLAC National
Accelerator Laboratory, Menlo Park, California 94025, USA}
\affiliation{Department of Applied Physics, Stanford University, Stanford, California
94305, USA}
\affiliation{PULSE Institute of Ultrafast Energy Science, SLAC National Accelerator
Laboratory, Menlo Park, California 94025, USA}
\email{huangyj@stanford.edu}

\author{Shan Yang}%
\affiliation{Department of Mechanical Engineering and Materials Science, Duke University, 
Durham, North Carolina 27708, USA }

\author{Samuel Teitelbaum}%
\affiliation{Stanford Institute for Materials and Energy Sciences, SLAC National
Accelerator Laboratory, Menlo Park, California 94025, USA}
\affiliation{PULSE Institute of Ultrafast Energy Science, SLAC National Accelerator
Laboratory, Menlo Park, California 94025, USA}
\thanks{Present Address: Department of Physics, Arizona State University, Tempe, AZ 85287, USA}

\author{Gilberto De la Pe\~na}
\affiliation{Stanford Institute for Materials and Energy Sciences, SLAC National
Accelerator Laboratory, Menlo Park, California 94025, USA}
\affiliation{PULSE Institute of Ultrafast Energy Science, SLAC National Accelerator
Laboratory, Menlo Park, California 94025, USA}

\author{Takahiro Sato}
\affiliation{Linac Coherent Light Source, SLAC National Accelerator Laboratory, Menlo Park, California
94025, USA}

\author{Matthieu Chollet}
\affiliation{Linac Coherent Light Source, SLAC National Accelerator Laboratory, Menlo Park, California
94025, USA}

\author{Diling Zhu}
\affiliation{Linac Coherent Light Source, SLAC National Accelerator Laboratory, Menlo Park, California 94025, USA}

\author{Jennifer L. Niedziela}
\affiliation{Department of Mechanical Engineering and Materials Science, Duke University, 
Durham, North Carolina 27708, USA}
\affiliation{Materials Science and Technology Division, Oak Ridge National Laboratory, Oak Ridge, Tennessee 37831, USA}

\author{Dipanshu Bansal}
\affiliation{Department of Mechanical Engineering and Materials Science, Duke University, 
Durham, North Carolina 27708, USA}

\author{Andrew F. May}
\affiliation{Materials Science and Technology Division, Oak Ridge National Laboratory, Oak Ridge, Tennessee 37831, USA}

\author{Aaron M. Lindenberg}
\affiliation{Stanford Institute for Materials and Energy Sciences, SLAC National
Accelerator Laboratory, Menlo Park, California 94025, USA}
\affiliation{PULSE Institute of Ultrafast Energy Science, SLAC National Accelerator
Laboratory, Menlo Park, California 94025, USA}

\affiliation{Department of Materials Science and Engineering, Stanford University, Stanford, CA 94305, USA}

\author{Olivier Delaire}
\affiliation{Department of Mechanical Engineering and Materials Science, Duke University, 
Durham, North Carolina 27708, USA }
\affiliation{Department of Physics, Duke University, 
Durham, North Carolina 27708, USA }
\affiliation{Department of Chemistry, Duke University, 
Durham, North Carolina 27708, USA }

\author{David A. Reis}
\affiliation{Stanford Institute for Materials and Energy Sciences, SLAC National
Accelerator Laboratory, Menlo Park, California 94025, USA}

\affiliation{Department of Applied Physics, Stanford University, Stanford, California
94305, USA}
\affiliation{PULSE Institute of Ultrafast Energy Science, SLAC National Accelerator
Laboratory, Menlo Park, California 94025, USA}
\affiliation{Department of Photon Science, Stanford University, Stanford, California
94305, USA}

\author{Mariano Trigo}
\affiliation{Stanford Institute for Materials and Energy Sciences, SLAC National
Accelerator Laboratory, Menlo Park, California 94025, USA}
\affiliation{PULSE Institute of Ultrafast Energy Science, SLAC National Accelerator
Laboratory, Menlo Park, California 94025, USA}

\begin{abstract}
There is growing interest in using ultrafast light pulses to drive functional materials into nonequilibrium states with novel properties. The conventional wisdom is that above gap photoexcitation behaves similarly to raising the electronic temperature and lacks the desired selectivity in the final state.  
Here we report a novel nonthermal lattice instability induced by ultrafast above-gap excitation in SnSe, a representative of the IV-VI class of semiconductors that provides a rich platform for tuning material functionality with ultrafast pulses due to their multiple lattice instabilities.
The new lattice instability is accompanied by a drastic softening of the lowest frequency A$_g$ phonon. This mode has previously been identified as the soft mode in the thermally driven phase transition to a Cmcm structure. However, by a quantitative reconstruction of the atomic displacements from time-resolved x-ray diffraction for multiple Bragg peaks and excitation densities, we show that ultrafast photoexcitation with near-infrared (1.55 eV) light, induces a distortion towards a different structure with Immm symmetry. The Immm structure of SnSe is an orthorhombic distortion of the rocksalt structure and does not occur in equilibrium.
Density functional theory (DFT) calculations reveal that the photoinduced Immm lattice instability arises from electron excitation from the Se 4$p$- and Sn 5$s$-derived bands deep below the Fermi level that cannot be excited thermally.
The results have implications for optical control of the thermoelectric, ferroelectric and topological properties of the monochalcogenides and related materials. More generally, the results emphasize the need for ultrafast structural probes to reveal distinct atomic-scale dynamics that are otherwise too subtle or invisible in conventional spectroscopies.

\end{abstract}

\maketitle

  \begin{figure*}
	\centering
	\includegraphics{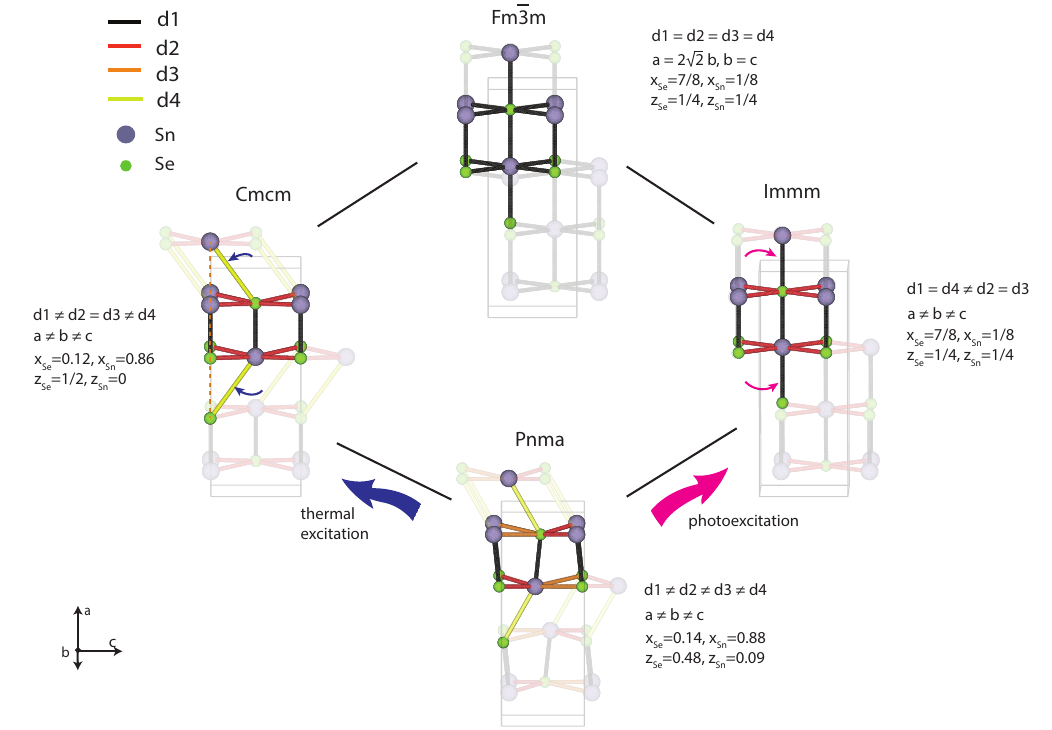}
	\caption{Relations between local coordination and atomic positions for different SnSe structures. At ambient conditions, the Pnma structure has the Sn and Se atoms off-center in the $\bf{b-c}$ plane, and is heavily distorted from the symmetric parent cubic structure Fm$\bar{3}$m ($d_1 =d_4 = d_2 = d_3 $). Bonds of the same color are equivalent under the symmetry of the given lattice. Above 807~K~\citep{li2015orbitally}, SnSe stabilizes in Cmcm ($d_1\neq d_2 = d_3 \neq d_4$) where $d_4$ bonds rotate further away from the parent rocksalt structure compared to Pnma. Orange broken lines in the Cmcm structure highlight the atoms located in the same $\bf{a-b}$ plane. Under photoexcitation, the atoms move towards the Immm structure, which is the highest symmetry orthorhombic distortion of the rocksalt structure ($d_1 =d_4\neq d_2 = d_3 $). We parameterize all crystal structures using the orthorhombic Pnma primitive unit cell as shown here (see also Appendix A Table~\ref{tab:Wyckoff Site table} and Fig.~\ref{fig:symmetry unit cell}), unless otherwise stated. The atomic positions are specified by $\pm(x_s,\frac{1}{4},z_s)$ and $\pm(\bar{x}_s+\frac{1}{2},\frac{3}{4},z_s+\frac{1}{2})$($s\in\{ \text{Sn}, \text{Se} \}$), where $x_s$ and $z_s$ are the fractional coordinates along $\bf{a}$ and $\bf{c}$ axis. Atoms in both the Immm and Fm$\bar{3}$m structures are located at high symmetry positions $x_{\text{Se}}=\frac{7}{8}, z_{\text{Se}}=\frac{1}{4}, x_{\text{Sn}}=\frac{1}{8}, z_{\text{Sn}}=\frac{1}{4}$. In the Cmcm phase atoms occupy different high symmetry positions $z_{\text{Sn}}=0, z_{\text{Se}}=\frac{1}{2}$. The Pnma fractional coordinates are taken from DFT calculations (see Appendix C), and the Cmcm fractional coordinates are taken from reference ~\cite{sist2016crystal}. Visualization made with VESTA~\citep{momma2008vesta}.}  
	\label{fig:symmetry unit cell_maintext}
\end{figure*} 
Ultrafast photoexcitation can alter the delicate energetic balance between nearly-degenerate material phases and the energy barriers separating them, potentially producing structures with novel functional properties not accessible in thermal equilibrium~\citep{basov2017towards}. Unlike in molecular systems, where coherence in electronic and vibrational degrees of freedom may be exploited to effect different reaction pathways using tailored light excitation\citep{Nuernberger2007Femtosecond}, 
in solids, it is often sufficient to assume that following above-gap excitation, electrons and holes quickly relax, resulting in a quasi-equilibrium electronic distribution that largely loses memory of the initial excitation\citep{anisimov1974electron,qiu1993heat,Shin2015}, ostensibly limiting the prospects of materials control. 
Indeed it is observed that in many ultrafast materials transformations it is sufficient to treat the photoinduced electronic excitation as an effective parameter similar to temperature within a time-dependent Ginzburg Landau model to describe the subsequent dynamics (for example ~\citep{Huber2014,Beaud2014a,trigo2021}).
Thus, there has been significant effort in using alternative ways to excite structural distortions in materials below-band-gap (e.g. terahertz and mid-infrared fields) in an attempt to avoid heating the electronic degrees of freedom~\citep{rini2007control,fausti2011light,mankowsky2014nonlinear,Li2019,kozina2019terahertz,Kubacka1333,kampfrath2013resonant}.
Although it is widely accepted that effective temperature models cannot be entirely correct, deviations often appear too subtle or are invisible to conventional spectroscopies. 
Here we use ultrafast diffraction and first-principle calculations to show that photoexcitation of SnSe, a prototypical functional material, results in a structural instability that is distinct from that achieved by raising the temperature (Fig. \ref{fig:symmetry unit cell_maintext}).  
The experiment is enabled by recent advances in free electron laser sources\citep{Emma2010first,Ishikawa2012compact,Abeghyan2019,Kang2017hard,milne2017swissfel}, 
which allow microscopic understanding of ultrafast materials dynamics~\citep{Wall2016recent,Lindenberg2017,Buzzi2018probing,Dunne2018x-ray,Cao2019ultrafast}.
The analysis is unambiguous based on a quantitative reconstruction of the sudden structural distortion and accompanying coherent phonon motion, including their \emph{phases}, following photoexcitation. The constrained ab-initio calculations help us identify the electronic states primarily involved in the observed new lattice instability. The results suggest that a better understanding of the initial electronic excitations and probing of the coupled atomic motions may enable a more microscopic approach to materials control with above-gap excitation.

\begin{figure*}
\centering
\includegraphics{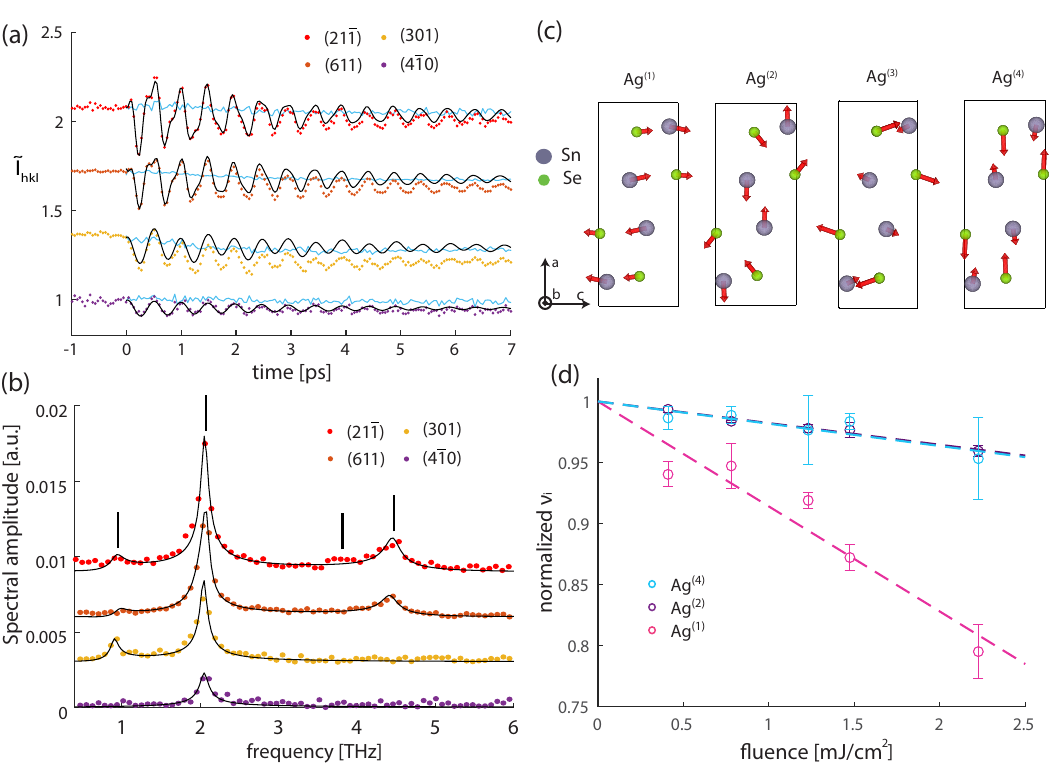}
\caption{\label{fig:raw data} (a) Normalized diffraction intensity $\tilde{I}_{hkl}(t)$ for $hkl = (4\bar{1}0)$, $(301)$, $(611)$, $(21\bar{1})$ Bragg peaks at a nominal absorbed fluence of $0.8~\mathrm{mJ/cm^2}$. Filled circles: experimental data. Black lines: decomposition of the data in the form of Eq.~(\ref{eq:intensity of DECP decomposition}). Traces are offset vertically for clarity. Blue lines: data subtracted by the most significant components. (b) Fourier transform (colored dots) of the data in (a), and sum of Lorentzians with frequency $\nu_i$, damping $\gamma_i$, phase $\phi_i$ and amplitude $B_{hkl}^{(i)}$ retrieved from the decomposition (black line). Three out of the four Raman active A$_g$ modes in Pnma SnSe are observed in the time resolved x-ray scattering data. The frequencies of the A$_g$ modes measured in Raman spectroscopy (see Appendix B) are indicated with short black bars. (c) Eigendisplacements ($\times$ 30) of the four $A_g$ modes. (d) Normalized mode frequencies as a function of nominal absorbed fluence from the analysis of the (21$\bar{1}$) Bragg peak.}
\end{figure*}
    
SnSe is a representative of rocksalt-like IV-VI compounds that hosts a number of lattice instabilities associated with their nearly cubic resonant bonding network. Differences in ionicity and spin-orbit coupling control the orbital hybridizations and lead to a diverse range of structural phases~\citep{cohen1964crystal,Littlewood1980a, Behnia2016}. The stability of these phases is sensitive to external parameters including temperature, pressure~\citep{chattopadhyay1986neutron}, as well as stoichiometry~\citep{lencer2008map}, stemming from the large polarizability that has its origin in unsaturated resonant bonding~\citep{lee2014resonant} and electron phonon interactions~\citep{jiang2016origin}. The large polarizability gives rise to a strong lattice anharmonicity, which leads to multiple fundamental and technologically relevant functionalities such as exceptional thermoelectric performance~\citep{heremans2008enhancement,Pei2011high,zhao2014ultralow,zhao2016ultrahigh,Chang2018,zheng2018rhombohedral,Aseginolaza2019,jiang2021high}, phase change behavior~\citep{lencer2008map,wuttig2018incipient}, ferroelectricity in 2D layers~\citep{wu2016intrinsic,wang2017two,xiao2018elemental,xu2021interlayer}, and antiferroelectricity in the bulk. Furthermore, SnSe~\citep{jin2017electronic}, as well as the related tertiary compound Pb$_{1-x}$Sn$_{x}$Se~\citep{xu2012observation} and some other IV-VI semiconductors, was observed to be topological crystalline insulators (TCI)~\citep{hsieh2012topological,dziawa2012topological} in their rocksalt phases. The richness in structural phases make IV-VI compounds an ideal playground for optical manipulation of materials, which could inspire novel functionality by accessing new hidden structures. 

Under ambient conditions, SnSe stabilizes in a centrosymmetric layered orthorhombic Pnma structure~\citep{sist2016crystal,wiedemeier1979thermal}. Compared to the rocksalt parent structure, the Pnma phase breaks the symmetry between the six nearest neighbor bonds that connect atoms of different elemental species, featuring alternate shearing of the bilayers accompanied by the lengthening and rotation of $d_1$ and $d_4$, as well as buckling and off-centering in the $\bf{b-c}$ plane network formed by $d_2$ and $d_3$.
SnSe undergoes a second-order phase transition~\citep{li2015orbitally} at high temperature (807~K)~\citep{chattopadhyay1986neutron,li2015orbitally} or pressure (10.5~GPa)~\citep{loa2015structural} to a higher symmetry orthorhombic phase with space group Cmcm, where further shearing of the bilayers causes the Sn and Se atoms to align in the $\bf{a-b}$ plane as indicated with dashed orange line in Fig.~\ref{fig:symmetry unit cell_maintext}.
In this work we find that photoexcitation induces an instability towards a new structure that is distinct from this high temperature Cmcm phase. As our analysis below shows, this lattice instability is towards an Immm structure, the highest symmetry orthorhombic distortion of rocksalt, where there is no off-centering and the bilayers are not sheared ($d_1=d_4$ and $d_2 = d_3$ as shown in Fig. \ref{fig:symmetry unit cell_maintext}). Importantly, Cmcm and Immm are not of a group-subgroup relationship.

The experiment was performed at the x-ray pump probe (XPP) end-station at the Linac Coherent Light Source (LCLS) x-ray free-electron laser (FEL). The near infrared (NIR) pump pulses  with photon energy 1.55~eV were derived from a Ti:sapphire laser, and the x-ray probe pulses had a photon energy of 9.5~keV~\citep{chollet2015x,zhu2015phonon}. The x rays were monochromatized using a diamond (111) double crystal monochromator, providing a nominal flux of $> 10^9$ photons per pulse. The sample is a single crystal SnSe grown with a Bridgman-type technique~\citep{li2015orbitally} and was polished with [100] surface normal. The x rays illuminated the sample at nominal grazing incidence of 0.5$^{\circ}$ with respect to the sample surface to match the penetration depth with the NIR laser, while the NIR beam was nearly colinear at an nominal incident angle of 1$^{\circ}$. A fast scan delay stage controlled the nominal delay between the NIR and x-ray pulses. Scattered x rays were collected by the Cornell-SLAC pixel array detector (CSPAD)~\citep{hart2012cspad}. Both x-ray and NIR pulses were $< 50$~fs. The relative arrival time $t$ between the x-ray probe and NIR pump was obtained on a shot-by-shot basis~\citep{harmand2013achieving}, and the x-ray scattering images were binned with intervals of 33~fs based on the sorted $t$. Multiple x-ray diffraction peaks were accessed by rotating the sample about the sample surface normal (azimuth), at nominally fixed grazing incidence. 

Fig.~\ref{fig:raw data}~(a) shows $\tilde{I}_{hkl} = I_{hkl}(t)/I_{hkl}(t<0)$ where $I_{hkl}(t)$ is the integrated intensity of the Bragg peak for $hkl=(21\bar{1})$, $(611)$, $(301)$ and $(4\bar{1}0)$~\footnote{$t<0$ corresponds to the x-ray probe pulse arriving before the NIR pump. We also note that $I(t<0)$ is virtually identical to the diffraction intensity without the pump.}. The data were taken at a nominal absorbed fluence of $0.8 \mathrm{mJ/cm^2}$.
The traces are offset vertically for clarity. We observe a combination of coherent oscillations in time, the frequency of which are associated with three of the four A$_{g}$ Raman active modes of the Pnma structure~\citep{chandrasekhar1977infrared}. 
Assuming the normalized intensity of the Bragg peaks $\tilde{I}_{hkl}(t>0)$ are composed of damped harmonic oscillators, we use linear prediction to decompose the time domain data ~\citep{barkhuijsen1985retrieval} and obtain robust, highly reproducible oscillator parameters (See Appendix D). 
The normalized intensity of each individual Bragg peak is well described by a sum of decaying cosines such that:
\begin{equation} \label{eq:intensity of DECP decomposition}
	\tilde{I}_{hkl}(t)=1+\sum_{i}B_{hkl}^{(i)}(1-e^{-\gamma_i t}\cos(2\pi \nu_i t+\phi_i)), 
\end{equation} 
with up to three components $i$. 
We absorb the sign of the initial intensity change into $B_{hkl}^{(i)}$, and find $\phi_i$ to be within $\pm 0.1\pi$. The small absolute value of $\phi_i$ is consistent with a macroscopic atomic motion produced via displacive excitation of coherent phonons (DECP) induced by above-gap excitation~\citep{zeiger1992theory} and the ensuing interatomic force changes~\citep{teitelbaumBismuthSoftening2020}. DECP typically involves A$_{g}$ phonons~\citep{zeiger1992theory,merlin1997generating} which fully respect the symmetry of the initial state and can potentially connect to higher symmetry phases via displacive phase transitions~\citep{zeiger1992theory}. The decomposition of the experimental data in the form of Eq.~(\ref{eq:intensity of DECP decomposition}) is shown with black lines in Fig.~\ref{fig:raw data} (a)  for a nominal incident fluence of 0.8 mJ$/$cm$^2$. Inclusion of only three oscillators captures well the observed time dependence. The residuals (blue lines in Fig.~\ref{fig:raw data} (a)) show a slowly varying, non-exponential background. %not accounted for in Eq.~(\ref{eq:intensity of DECP decomposition}). 
This slowly varying background is likely due to strain that develops and propagates over the probed volume on a much longer timescale than the optical phonons~\citep{reis2006ultrafast} and thus the relevant materials dynamics can be considered as occurring at a constant volume.

The colored dots in Fig.~\ref{fig:raw data} (b) show the magnitude of the Fourier transform of the data in Fig.~\ref{fig:raw data} (a). The black lines in (b) show the sum of Lorentzians obtained from the fitted frequency, amplitude and damping constant in Eq.(\ref{eq:intensity of DECP decomposition})\footnote{The spectral amplitude of $\tilde{I}_{hkl}$ is $\Big| \sum_i B_{hkl}^{(i)} \frac{-(2\pi\nu_i)\sin{\phi_i}+(\gamma_i-2\pi i \nu_i)\cos{\phi_i}}{(2\pi\nu_i)^2+(\gamma_i-2\pi i \nu_i)^2}\Big|$}. The data clearly reveal three modes at 0.9~THz, 2 THz and 4.5 THz, in agreement with the frequencies of the A$_g^{(1)}$, A$_g^{(2)}$ and A$_g^{(4)}$ modes observed in Raman measurements~\citep{chandrasekhar1977infrared}. 
Although the A$_g^{(3)}$ mode is visible in ultrafast pump-probe optical reflectivity as well as Raman scattering measurements (see Appendix B), the x-ray data does not show a strong signature of the A$_g^{(3)}$ above the noise (see residual traces in light blue in Fig.~\ref{fig:raw data}~(a)). This is likely due to a combination of relatively short A$_g^{(3)}$ lifetime and a limited sensitivity to A$_g^{(3)}$ motion for the measured Bragg peaks. The eigendisplacements ($\times$ 30) of the four A$_g$ modes obtained from harmonic phonon calculation with DFT are shown in Fig.~\ref{fig:raw data}~(c)~\footnote{From DFT we obtain interatomic force constants and recast them into the dynamical matrix $D(\bf{k})$~\citep{born1954dynamical,ashcroft1976solid}. The solutions to $D({\bf{k}}) \bm{\epsilon}_{{\bf{k}}}^{(i)}=\omega_{i}^{2}({\bf{k}})\bm{\epsilon}^{(i)}_{{\bf{k}}}$ yield the orthonormal eigenvectors $\bm{\epsilon}_{{\bf{k}}}^{(i)}$, where ${\bf{k}}$ represents the wave vector. For a SnSe conventional unit cell that contains 8 atoms, the eigenvector $\bm{\epsilon}_{{\bf{k}}}^{(i)}$ should contain 8 $\times$ 3 = 24 elements. We only discuss zone center phonon (${\bf{k=0}}$) and simplify the notation $\bm{\epsilon}_{{\bf{k=0}}}^{(i)}$ to $\bm{\epsilon}^{(i)}$. The A$_g^{(i)}$ displacement of the $\sigma$ atom  ${\bf{u}}_{\sigma}$, 
\begin{equation} 
\begin{split}
    {\bf{u}}_{\sigma}^{(i)}=\text{Re}\Big\{ \alpha_i \frac{1}{\sqrt{\mu_\sigma}} {\bm{\epsilon}}^{(i)}_{\sigma}\mathrm{exp}[-i\omega_i t]\Big\}
\end{split}
\end{equation}
is proportional to the reduced amplitudes $\alpha_i$ which are scalars, and the eigendisplacement, which is the eigenvector normalized by square root of atomic masses. ${\bm{\epsilon}}^{(i)}_{\sigma}$ is the atomic projection of the eigenvector $\bm{\epsilon}^{(i)}$, and is a (3 $\times$ 1) vector, as is ${\bf{u}}_{\sigma}^{(i)}$. }. 
The A$_g^{(1)}$ displacements mainly involve shearing between bilayers along the $\bf{c}$-axis, while A$_g^{(2)}$ mainly affects the buckling the bilayer by moving Sn and Se from the same atomic layer oppositely along $\bf{a}$-axis. The A$_g^{(3)}$ and A$_g^{(4)}$ displacements are similar to those of the A$_g^{(1)}$ and A$_g^{(2)}$ modes respectively, but with Sn atoms $\pi$ out of phase. Note that in DECP, depending on the initial phase of motion, the atomic motion direction can be opposite to what is shown in Fig.~\ref{fig:raw data}~(c). 
\begin{figure}
\includegraphics{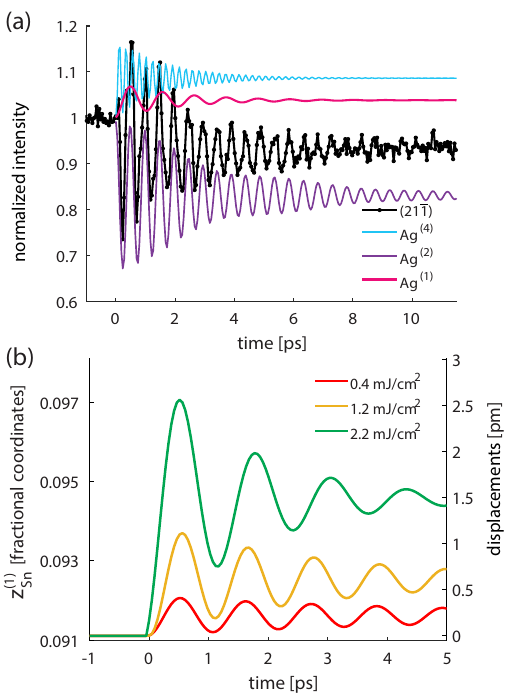}
\caption{\label{fig:Ag1 mode} (a) Decomposition of the $(21\bar{1})$ time-resolved diffraction signal at a nominal absorbed fluence of $0.8 \mathrm{mJ/cm^2}$ according to Eq. (\ref{eq:intensity of DECP decomposition}). (b) Time dependence of $z_{\rm Sn}^{\text(1)}$, the Sn $z$ position projected onto the A$_g^{(1)}$ mode, as calculated from the $(21\bar{1})$ peak measured under different nominal absorbed fluences. $z_{\rm Sn}^{\text(1)}$ is defined in fractional coordinates. 
The right y-axis shows the corresponding displacements in picometers (pm). }
\end{figure}

The A$_g^{(1)}$ phonon mode is of particular interest because it becomes unstable as the temperature approaches the Pnma-Cmcm phase transition at $T_c = 807$~K~\citep{li2015orbitally}, and it strongly overlaps with the order parameter of this second order phase transition\citep{chattopadhyay1986neutron,li2015orbitally,hong2019phase,Lanigan2020}. At $T > T_c$ this mode becomes an acoustic mode at the zone boundary $Y$ of the Cmcm Brillouin zone. Fig. \ref{fig:raw data} (d) shows the normalized mode frequency as a function of nominal absorbed fluence. The y-axis is obtained by normalizing the fitted frequency $\nu_i~(i\in \{1,2,4\} )$ from Eq.~(\ref{eq:intensity of DECP decomposition}) by their zero-fluence extrapolations. The A$_g^{(1)}$ mode softens (decreases frequency) as much as 20\% at the highest nominal absorbed fluence of the experiment (2.2~mJ/cm$^2$), while the A$_g^{(2)}$ and A$_g^{(4)}$ modes soften less than 4\%. Error bars of the frequencies represent the statistical error due to the shot-to-shot FEL intensity fluctuations and are estimated using the standard error of the $\nu_i$ ensemble obtained by the decomposition of random sub-samples of the data. The strong softening of $\nu_1$ suggests the existence of a lattice instability, i.e. a softening of the interatomic potential associated with the mode coordinate. Based on a robust analysis of $\tilde{I}_{hkl}(t)$, we show next that this instability is not associated with the thermal transition to Cmcm, but rather towards a structure that would resemble Immm for large displacements. 

In the kinematic diffraction limit, the intensity of the $(hkl)$ Bragg peak using the Pnma unit cell convention, is  
\begin{equation} \label{eq:structure factor}
\begin{split}
    \tilde{I}_{hkl}(t) \propto \Big|\sum_{{s\in {\rm Se,Sn}}}  4f_s \cos \left[ 2\pi \left(   h x_s(t) + \frac{h+k+l}{4} \right)\right] \times\\
    \cos \left[2\pi\left(l z_s(t) - \frac{h+l}{4}\right)\right] \Big|^2,
    \end{split}
\end{equation}
where $f_s$ are the atomic form factors. The four parameters $x_{s}$, $z_{s}$($s\in\{ \text{Sn}, \text{Se} \}$) fully represent the atomic positions of the eight atoms in a Pnma unit cell in fractional coordinates $\pm(x_s,\frac{1}{4},z_s)$ and $\pm(\bar{x}_s+\frac{1}{2},\frac{3}{4},z_s+\frac{1}{2})$\citep{Wyckoff}. 
According to Eq~(\ref{eq:structure factor}), the intensity of Bragg peaks where $h+l=\mathrm{odd}$ (e.g. (21$\bar{1}$) and (611)) decreases monotonically to zero as $z_{\rm{Sn}}\to 0$ and $ z_{\rm{Se}}\to \frac{1}{2}$ in the Cmcm phase. 
However, as shown in Fig.~\ref{fig:Ag1 mode}~(a), where we show the mode decomposition of the (21$\bar{1}$) data as an example, the A$_g^{(1)}$ component (pink trace) oscillates around an increased intensity ($\tilde{I}_{hkl}>1$), indicating that $z^{(1)}_{\rm Sn}$, the A$_g^{(1)}$ mode-projected $z_{\rm Sn}$ motion, moves \emph{away} from, rather than towards zero. See Fig.~\ref{fig:Ag1 mode} (b) for $z^{(1)}_{\rm Sn}$ measured under several fluences on Bragg peak (21$\bar{1}$). As a supplement, the mode decompositions featuring A$_g^{(1)}$ are displayed in Appendix D for other Bragg peaks. Since the A$_g^{(1)}$ involves primarily $z_{\rm Sn}$ motion, and it strongly overlaps with the order parameter of Pnma-Cmcm phase transition, photoexcitation of the A$_g^{(1)}$ mode appears to displace the lattice further away from the Cmcm structure.

To further refine this observation we quantitatively reconstruct the photoexcited atomic motion using the diffraction data. 
We will show below in Fig.~\ref{fig:reconstruction} (a) that the observed $\Delta z^{(1)}_{\rm Sn}>0$ which we illustrate using Bragg peak (21$\bar{1}$) as an example in Fig.~\ref{fig:Ag1 mode}, is consistent for all measured Bragg peaks and excitation fluences.
We first use Eq.~(\ref{eq:intensity of DECP decomposition}) to obtain $B_{hkl}^{(i)}$ for each individual Bragg peak $(hkl)$ and each excitation fluence. Then based on Eq.~(\ref{eq:structure factor}), we use $B_{hkl}^{(i)}$ and the eigendisplacements,  shown in Fig.~\ref{fig:raw data} (c), to obtain a dimensionless amplitude $\alpha_i$ [67]. 
Fig.~\ref{fig:Ag1 mode} (b) shows the dynamics of $z_{\rm{Sn}}^{\text(1)}$, the A$_g^{(1)}$-projected $z_{\rm{Sn}}$ displacements, extracted from the (21$\bar{1}$) peak under nominal absorbed fluences 0.4, 1.2 and 2.2~mJ/cm$^2$. For reference, we also show the $z_{\mathrm{Sn}}^{(1)}$ displacement in picometers (pm) on the right y-axis in Fig.~\ref{fig:Ag1 mode} (b).  The A$_g^{(1)}$ motion increases $z_{\rm{Sn}}$ from the initial value $z_{\rm{Sn}}=0.09$, instead of decreasing it towards $z_{\rm{Sn}}=0$, the atomic position of the Cmcm phase. The fact that photoexcitation both softens the the A$_g^{(1)}$ mode and produces a shift in its quasi-equilibrium position \emph{further away} from Cmcm, signals an instability associated with a new transient structure. 

\begin{figure}
\includegraphics{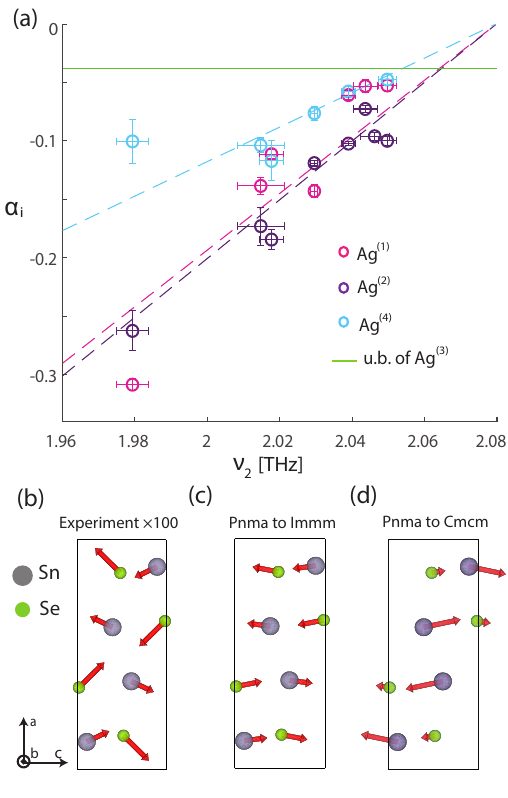}
\caption{\label{fig:reconstruction}
(a) $\alpha_i$ as a function of A$_g^{(2)}$ frequency $\nu_2$, and corresponding linear fits. Data points are obtained from the four Bragg peaks $(21\bar{1})$, $(611)$,$(301)$ and $(4\bar{1}0)$ under a range of excitation level. lower $\nu_2$ corresponds to higher absorbed fluence. Solid green line represents the upperbound of $|\alpha_3|$. (b) Photoexcition induced shift of atomic quasi-equilibrium in SnSe (red arrows) under a nominal absorbed fluence of 2.2~mJ/cm$^2$, magnified $\times 100$. (c), (d) Displacement connecting Pnma with Immm (c) and Cmcm (d). }
\end{figure}

The transient structure can be understood by combining all A$_g$ displacements and reconstructing the overall lattice distortion in the photoexcited state. To consistently incorporate \emph{all} data sets with a variety of pump fluences and Bragg peaks, we take the degree of A$_g^{(2)}$ frequency softening as a more accurate measure of the absorbed fluence than the nominal absorbed fluence, which is obtained from the measured pulse energy and the illuminated area~\footnote{This is to mitigate the effect of systematic errors due to the anisotropy of refractive index, which results in variations of optical absorption caused by azimuthal rotations. A$_g^{(2)}$ has good signal noise ratio in all the measured Bragg peaks.}. The reduced mode amplitudes $\alpha_i$ correlate linearly with A$_g^{(2)}$ mode frequencies as the absorbed fluence varies, as shown in Fig.~\ref{fig:reconstruction} (a). 
The dashed lines in Fig.~\ref{fig:reconstruction} (a) are fits of $\alpha_{i} = b_i(\nu_2 - \nu_2^{0})$, where $b_i$ are fitting parameters and the value of $\nu_{2}^{0}$ is fixed as the equilibrium A$_g^{(2)}$ frequency 2.08~THz, which is measured with Raman spectroscopy at room temperature (see Appendix B). The fit described above weighs in the error bars of both $\nu_2$ and $\alpha_i$~\footnote{The error bars for $\nu_2$ take into account only the statistical error, as in Fig.~\ref{fig:raw data} (d). The error bars for $\alpha_i$ take into account both the statistical error due to FEL intensity fluctuations, and the systematic error. The statistical error for $\nu_i$ ($B_{hkl}^{(i)}$), as described for Fig.~\ref{fig:raw data} (d), are estimated with the standard error of the $\nu_i$ ($B_{hkl}^{(i)}$) ensemble obtained by fitting Eq.~(\ref{eq:intensity of DECP decomposition}) to multiple random sub-samples of the data. The uncertainty of $B_{hkl}^{(i)}$ can then map to the uncertainty of $\alpha_i$. The systematic error is mainly attributed to a deviation of the grazing angle $\beta$ for different azimuthal rotations. This causes slight variation in the probed volume which leads to changes in the observed signal magnitude. The $\beta$ dependence of x-ray penetration depth is calculated based on~\citep{henke1993x}, and the effective $\beta$ is estimated to have a $\pm 0.05^{\circ}$ uncertainty due to sample surface flatness. To weigh in the error bars on both x and y axis, one can reference for example \citep{fiterror}}. 
We neglect A$_g^{(3)}$ for the reconstruction of the atomic motion because its amplitude upper bound (green line) is significantly smaller than other modes~\footnote{The upper bound of the A$_g^{(3)}$ mode amplitude $\alpha_3$ is estimated from the noise level in the ($21\bar{1}$) Bragg peak data set shown in Fig.~\ref{fig:raw data} (a), taking into  account the peak sensitivity (which is small but nonzero) to $\alpha_3$. The noise level is estimated from the larger one of the R.M.S of the $t > 0$ fit residual (light blue trace in Fig.~\ref{fig:raw data} (a)) and the R.M.S of the $t < 0$ pre-pump noise.}.
Summing the product of $\alpha_i$ and mode eigendisplacement (Fig.\ref{fig:raw data} (c)) for all the phonon modes, one obtains the overall displacements of the quasi-equilibrium atomic positions, described by $\Delta x_s$ and $\Delta z_s$. 
We plot in Fig. \ref{fig:reconstruction} (b) this reconstructed overall atomic displacement ($\times 100$).  $\alpha_i$ used in the reconstruction is predicted by the linear fit in Fig. \ref{fig:reconstruction} (a) at the minimum observed $\nu_2$ value, i.e., the maximum absorbed fluence in the experiment, which is nominally 2.2$\mathrm{mJ/cm^2}$). 
To be specific, Fig. \ref{fig:reconstruction} (b) shows
$\Delta x_{\rm Se}=(1.2\pm 0.1) \times 10^{-3}$, $\Delta z_{\rm Se}=(-2.9\pm 0.3) \times 10^{-3}$, $\Delta x_{\rm Sn}=(2.6\pm 1.6)\times 10^{-4}$ and $\Delta z_{\rm Sn}=(2.0\pm 0.3) \times 10^{-3}$~\footnote{These displacements in absolute units are $a\Delta x_{\rm Se}=1.34\pm0.15$~pm, $c\Delta z_{\rm Se}=-1.25\pm0.11$~pm, $a\Delta x_{\rm Sn}=0.29\pm0.17$~pm and $c\Delta z_{\rm Sn}=0.85\pm0.13$~pm. }. The signs of $\Delta x_{s}$ and $\Delta z_{s}$ are robust within the experimental uncertainties.

To identify a new lattice structure that is compatible with a large amplitude extrapolation of the displacements presented above, it is instructive to search for higher symmetry space groups (supergroups of Pnma), since a displacement along a linear combination of A$_g$ modes cannot lower the lattice symmetry. We restrict the search among orthorhombic space groups since as stated earlier we approximate the lattice constants to be fixed on the few-ps timescale.
The A$_g$ displacement connecting Pnma to the new structure must be consistent with the experimental observation $\Delta x_{\rm Se}>0$, $\Delta z_{\rm Se}<0$, $\Delta x_{\rm Sn}>0$, $\Delta z_{\rm Sn}>0$. 
Based on these criteria we identify Immm as the space group associated with the photoexcited lattice instability (see Appendix A). This conclusion is independent of the exact numerical values of $\Delta x_{s}$ and $\Delta z_{s}$. The atomic displacements connecting Pnma to Immm (Cmcm) structures are plotted in Fig.~\ref{fig:reconstruction}~(c) ((d)) to scale. Clearly, the signs of $\Delta x_{s}, \Delta z_{s}$ rule out a distortion towards Cmcm as has been shown in Fig~\ref{fig:Ag1 mode} (b).
The magnitudes of $\alpha_1$ and $\alpha_2$ are significantly larger than $\alpha_3$ and $\alpha_4$ (Fig.~\ref{fig:reconstruction} (a)). This is consistent with the fact that atomic displacements associated with the Pnma-Immm lattice instability (Fig.~\ref{fig:reconstruction} (c)) can be decomposed into restorations of high symmetry positions along the $\bf{c}$- and $\bf{a}$-axes, dominated by motion along the A$_g^{(1)}$ and A$_g^{(2)}$ coordinates respectively. In both the transient photoexcited (Fig.~\ref{fig:reconstruction} (b)) and the Immm structures (Fig.~\ref{fig:reconstruction} (c)), the component of the displacements along $\mathbf{c}$ relative to the Pnma structure mainly involves the A$_g^{(1)}$ mode, which provides the inter-layer shearing towards a rocksalt-like stacking and tends to align Sn and Se atoms alternately along $\bf{a}$-axis with high symmetry positions $z_{\rm Sn}=\frac{1}{4}, z_{\rm Se}=\frac{1}{4}$. Whereas the  component of the reconstructed motion along $\mathbf{a}$ mainly involves the A$_g^{(2)}$ mode, which reduces the buckling of the bilayers in the $\mathbf{a}$-axis and brings the atoms closer to the high symmetry positions of the Immm structure $x_{\rm Sn}=\frac{1}{8}, x_{\rm Se}=\frac{7}{8}$. 
The total photoexcited atomic displacements reduce the difference between the $d_1$ and $d_4$ bond lengths, consistent with the Immm structure but inconsistent with the Cmcm structure (Fig.~\ref{fig:symmetry unit cell_maintext}). A detailed analysis of bond lengths and bond angles is provided in Appendix A. Our identification of the photoinduced lattice instability towards Immm instead of Cmcm is robust and is further supported by DFT calculations detailed below.

\begin{figure}
\centering 
\includegraphics{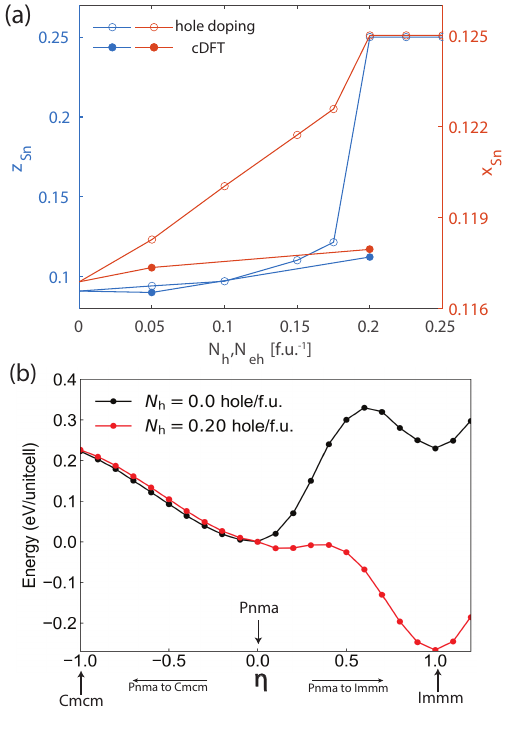}
\caption{(a) Computed $x_{\rm Sn}$, $z_{\rm Sn}$ as a function of hole concentration, $N_{h}$ (hole doped DFT, hole/f.u.) and electron-hole concentration, $N_{eh}$ (constrained DFT, pairs/f.u.), at constant volume. 
Both calculations show a tendency to distort towards the Immm structure with increasing $N$.
For hole doping, the Immm structure ($z_{\rm Sn}=0.25$, $x_{\rm Sn}=0.125$) is obtained at and above $N_{h}$=0.2~hole/f.u. (b) Potential energy as a function of atomic configurations interpolated between Cmcm-Pnma ($-1\le \eta\le0$) and Pnma-Immm ($0\le \eta \le1$), at different hole doping levels $N_{h}$=0, 0.2 hole/f.u. The Immm structure is stabilized at $N_{h}$=0.2 hole/f.u, evidenced by the energy minimum at $\eta=1$.
} 
\label{fig:Sn_displacement}
\end{figure}

To gain insight into the observed lattice instability we performed constrained-DFT (cDFT) calculations where we model the effect of photoexcitation by constraining the electron occupations using two different chemical potentials for electrons and holes~\citep{tangney2002density}, while keeping the lattice constants fixed. cDFT calculations were performed with constrained densities $N_{eh}=0.05$ and $N_{eh}=0.20$ electron-hole pairs per formula unit (pairs$/$f.u.). For reference, from the experimental parameters we estimate $N_{eh}=0.15$~pairs$/$f.u. at 2.2~$\mathrm{mJ/cm^2}$. These were estimated from the absorbed energy density per unit volume considering the reflectivity, optical penetration depth of 60~nm and the illuminated pump area of $2$~mm$^2$~\footnote{The refractive index was obtained  from~\citep{LandoltBornstein1998:sm_lbs_978-3-540-31360-1_851}. }. 
In Fig.~\ref{fig:Sn_displacement} (a), we show the calculated $x_{\rm Sn}$ and $z_{\rm Sn}$  for several constrained charge densities $N_{eh}$ (solid symbols in (a)). The cDFT calculations predict that Sn atoms displace towards the Immm structure instead of towards the Cmcm structure, i.e. $z_{\rm Sn}$ increases rather than decreases with increasing $N_{eh}$. The direction is consistent with the experimental results (Figs.~\ref{fig:Ag1 mode} (b) and \ref{fig:reconstruction} (b)). Quantitatively, the calculated displacement $\Delta z_{\rm Sn}=0.02$ is an order of magnitude larger than the measured displacement $\Delta z_{\rm Sn}=2.0 \times 10^{-3}$ for 0.15~pairs/f.u. 

\begin{figure*}
\centering
\includegraphics{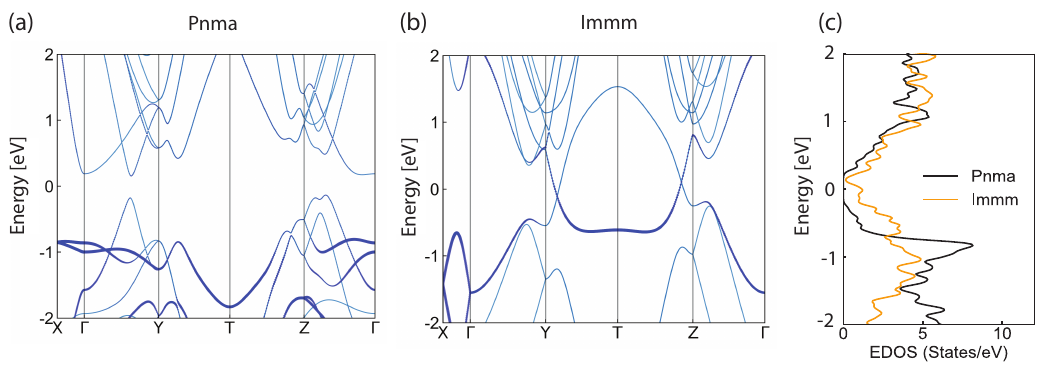}
\caption{ Calculated electronic band structure of (a) equilibrium Pnma phase ($N_{h}$ = 0.0\,hole/f.u.) and (b) hole-doped Immm structure ($N_{h}$=0.20\,hole/f.u.). The blue line thickness represents the band projection onto the Se 4$p_{x}$ orbital. (c) Electron density of states of Pnma ($N_{h}$ = 0.0\,hole/f.u.) and Immm ($N_{h}$=0.20\,hole/f.u.) structures.} 
\label{fig:electron_band_structure}
\end{figure*} 
To obtain further insight into the effect of photoexcitation and the corresponding energy landscape, we also performed a simplified version of cDFT calculations using hole doping, also at fixed lattice constants. Hole doping captures salient features of the distortion and has better numerical convergence than cDFT. As with cDFT, these calculations predict the structure of SnSe distorts from Pnma towards the higher symmetry Immm with increasing hole density. Hole doping DFT calculations predict an abrupt phase transition to Immm near 0.2\,hole/f.u., as shown in Fig.~\ref{fig:Sn_displacement} (a). In Fig.~\ref{fig:Sn_displacement} (b) we show the calculated total energy for a series of configurations between Pnma-Immm, and Pnma-Cmcm, at two different hole concentrations $N_{h}$ = 0 and 0.2\,hole/f.u. Here  $\eta$ parameterizes the structural configuration  representing  linear interpolations between Pnma ($\eta = 0$) and Immm ($\eta = +1$) and Cmcm  ($\eta = -1$). The abrupt structural phase transition upon hole doping at $N_h = 0.2$~hole/f.u, is accompanied by a suppression of the  0.33\,eV/unit-cell energy barrier (near $\eta=0.5$, for $N_{h}$ = 0\,hole/f.u). Meanwhile, the hole doping levels investigated ($N_{h}$ = 0, 0.2\,hole/f.u.) do not significantly affect the energy landscape between the Pnma and Cmcm structures. In particular, the energy of the Cmcm structure remains higher than that of Pnma when hole-doped. The energy minimum at $\eta=1$ under 0.2\,hole/f.u. suggests that the Immm structure may be realized at sufficiently high excitation density if sample damage can be mitigated. 

According to the hole doped DFT calculations, the photoinduced Immm structural instability could be attributed to the excitation of electrons out of Se 4$p_x$ orbital derived bands by the absorption of the NIR photons. Fig.~\ref{fig:electron_band_structure} shows the calculated electronic band structure of Pnma ($N_{h}$=0.0\,hole/f.u., Fig.~\ref{fig:Se p bands}(a)) and the transient photoexcited Immm ($N_{h}$=0.2\,hole/f.u., Fig.~\ref{fig:Se p bands}(b)) SnSe, respectively. The Brillouin zone labels follow the Pnma unit cell convention (see Fig~\ref{fig:BZ}). The thick portions of bands in Fig.~\ref{fig:electron_band_structure} (a) and (b) represent Se 4$p_x$ orbital character. 
In the Pnma phase, the Se 4$p_{y,z}$ orbitals hybridize with Sn 5$s$ orbitals and form a symmetric resonantly bonded network in the $\bf{b-c}$ plane. This hybridized orbital is the main character of the edge of the valence bands (see Appendix A Fig.~\ref{fig:Se p bands}) and contributes to the Pnma-Cmcm phase transition through a Peierls-like mechanism~\citep{hong2019phase,li2015orbitally}. 
The Se 4$p_x$ orbital, however, is hybridized with Sn 5$s$ (for Sn 5$s$ orbital projected band structure, see Appendix Fig.~\ref{fig:Se p bands}) to form the non-dispersive band along $\Gamma-X$ about 0.7~eV below the top of the valence band in Fig. \ref{fig:electron_band_structure} (a). From the electron density of states (EDOS) in Pnma phase (Fig.~\ref{fig:electron_band_structure} (c)), the holes will populate down to $-$0.77~eV under 0.2~hole/f.u assuming holes are filled from the top of the valence band, and $-$0.77~eV is close to the peak of the EDOS, which is mostly formed by the non-dispersive bands due to non-bonding lone pairs in Pnma (compare Fig.~\ref{fig:electron_band_structure} (a) and (c)). Removal of electrons from these non-dispersive bands causes suppression of the lone-pair stereo-chemical activity, which is generally considered ~\citep{orgel1959769} to raise the structural symmetry, in our case the symmetry of the local quasi-octahedral coordination.
%, reduces repulsion against $d_4$, among other nearest neighbor bonds, and favors restoring the symmetric local quasi-octahedral coordination that corresponds to Immm.  
In fact, in the Immm structure, the electron bands feature a clear dispersion of the band consisting mainly of Se 4$p_x$ and Sn 5$s$ orbital components (compare Fig. \ref{fig:electron_band_structure} (a) and (b) between $\Gamma-X$), which reflects the disappearance of the non-bonding localized lone-pair orbitals.

Experimentally we observe a significant lengthening of $d_1$ and concomitant shortening of $d_4$, \emph{opposite} the trend towards Cmcm (see Appendix A Table ~\ref{tab:bond length}). This is consistent with the weakening of $d_1$  and strengthening of $d_4$ forces as calculated in Appendix A Table ~\ref{tab:force constants} and can be explained by the change of intra-layer and inter-layer hybridization of the Sn 5$s$ and Se 4$p_x$ orbital-derived bands. 
The depopulation of the in-plane Se 4$p_{y,z}$ orbitals, however, is expected to strengthen the in-plane resonant bonds and soften the in-plane polarized A$_g^{(1)}$ modes in \emph{both} the Pnma-Cmcm~\citep{li2015orbitally} and  Pnma-Immm lattice instabilities. 
Apart from the two nearest neighbor resonant bonds ($d_2$ and $d_3$) that become equivalent in Immm, other resonant bonds connecting atoms distanced further apart all strengthen (Appendix C Fig.~\ref{fig:resonant bonds}), which is similar to the trends of force changes incurred under enhanced temperatures in the structural phase transition to Cmcm in both SnSe and the related material SnS~\citep{Lanigan2020}.
Clearly, the softening of A$_g^{(1)}$ alone cannot distinguish between a Pnma-Cmcm versus a Pnma-Immm phase transition in SnSe, highlighting the importance of ultrafast atomic-scale probes for resolving photoexcited atomic motion in materials with structural instabilities.

Fig.~\ref{fig:electron_band_structure}~(a) and (b) show that along $T-Y$ and $T-Z$ in Immm, the Se $p_x$ bands and Sn $p_x$ bands are inverted, and the band gaps that exist in Pnma close~\footnote{ $T-Y$ and $T-Z$ corresponds to $W-S$ and $W-R$ in Immm convention, see Fig.~\ref{fig:BZ}, which would correspond to  $W-L$ in the Fm$\bar{3}m$ Brillouin zone for the rocksalt structure.}. The band crossings along $T-Y$ and $T-Z$ do not occur at the same energy due to orthorhombic structural distortion, leading to a finite EDOS everywhere in Fig.~\ref{fig:electron_band_structure}. The Pnma-Immm structural instability and its connection to the disappearance of the lone pair is reminiscent of the structural phases formed by other group V or IV-VI rocksalt derived materials~\citep{tremel1987tin,GALY2020106068}.The photoinduced structural instability has a Peierls-mechanism nature, but orginates from different orbitals than the Pnma-Cmcm Peierls instability. 
Though the band inversion and gap closing along $T-Y$, $T-Z$ in SnSe Immm bear resemblance to the electron band dispersion in the rocksalt TCI  of IV-VI compounds ~\citep{hsieh2012topological,xu2012observation,sun2013rocksalt}, Immm lacks the proper lattice symmetry to become a TCI.

We note that DFT calculations assuming increased electronic temperature and the same chemical potential for both electrons and holes will leave the electrons occupying the lone-pair orbitals mostly intact, and does not even qualitatively reproduce the experimentally observed atomic motion (see Appendix C Fig.~\ref{fig:simulation}). This suggests that the theoretical formalism for non-equilibrium photoexcited material needs to be dealt with care in order to predict material behavior under above-gap excitation, and this effort will mostly likely need to be combined with microscopic experimental probes.

In summary, we have shown that ultrafast NIR photoexcitation of SnSe favors a structural instability towards Immm, an orthorhombically distorted rocksalt structure, rather than towards the thermodynamic Cmcm phase. Though both Cmcm-Pnma and the Immm-Pnma instabilities can be thought of as symmetry lowering due to a Peierls-like mechanism, they are related to different electron orbitals. Our DFT results suggest that the Immm instability is due to the excitation of electrons out of non-bonding Se 4$p_x$- Sn 5$s$ orbitals by the 1.55~eV laser pulse. Hence we demonstrated that nonequilibrium states induced by ultrafast light pulses can activate electron-phonon coupling mechanisms not manifested near thermal equilibrium. DFT calculations also suggest that at high hole-doping density, the Immm structure becomes stable. 
The experiments reported here were limited in the maximum carrier density we could produce without damaging the sample. 
We note that alloying with Pb may reduce the Immm phase-transition threshold~\citep{Littlewood1980a,inoue2016nonequilibrium,Behnia2016} while lowering the temperature will increase the damage threshold, such that a photoinduced Immm phase may be realizable. 
Our findings may also have implications in other rocksalt distorted IV-VI semiconductors, several of which have topological states protected by lattice symmetry in the cubic or tetragonal phases~\citep{hsieh2012topological,xu2012observation,dziawa2012topological,tanaka2012experimental}.
More generally, our work suggests that pump wavelength could provide additional control of structural distortions through orbitally-selective above-gap excitation. This could be exploited to direct a particular structural distortion to desirable outcomes with particular functionality beyond those accessible in thermal equilibrium.

\begin{acknowledgments}
Preliminary x-ray characterization was performed at beamline 7-2 at the Stanford Synchrotron Radiation Lightsource (SSRL). The Raman scattering measurement was performed at the Stanford Nano Shared Facilities (SNSF), supported by the National Science Foundation under award ECCS-2026822. Y. H., S.T., G.d.P, D.A.R. and M.T. were supported by the U.S. Department of Energy, Office of Science, Office of Basic Energy Sciences through the Division of Materials Sciences and Engineering under Contract No. DE-AC02-76SF00515. S.Y. acknowledges support by the Fitzpatrick Institute for Photonics through a Chambers Scholarship. O.D. acknowledges support from the U.S. Department of Energy, Office of Science, Basic Energy Sciences, Materials Sciences and Engineering Division, under Award No. DE-SC0019978. Use of the LCLS and SSRL is supported by the US Department of Energy, Office of Science, Office of Basic Energy Sciences under Contract No. DE-AC02-76SF00515. Sample synthesis and characterization (A.F.M.) was supported by the U. S. Department of Energy, Office of Science, Basic Energy Sciences, Materials Sciences and Engineering Division.

\end{acknowledgments}
\begin{table*}
		\caption {Wyckoff sites of Pnma SnSe, and the higher symmetry structures Cmcm, Pmmn, Immm. $x_s$ and $z_s$ are expressed in fractional coordnates\citep{Wyckoff,Wyckoffsplit}. } \label{tab:Wyckoff Site table} 
		\centering
		\begin{tabular}{c|c|c|c|c}
			\hline
			Pnma 4c &   Cmcm 4c& Pmmn 2b& Immm 2d(Se)&Immm 2b(Sn)  
			\\
			\hline
			$x_s$,1/4,$z_s$ 		& $x_s$,1/4,0		&	$x_s$,1/4,1/4		&-1/8,-3/4,1/4 & -7/8,-3/4,1/4
			\\
			1/2-$x_s$,3/4,1/2+$z_s$ & 1/2-$x_s$,3/4,1/2	&	1/2-$x_s$,3/4,3/4	&-3/8,-1/4,3/4 &-5/8,-1/4,3/4
			\\
			-$x_s$,3/4,-$z_s$ 		& -$x_s$,3/4,0 		&	-$x_s$,3/4,3/4		&-7/8,-1/4,3/4 &-1/8,-1/4,3/4
			\\
			1/2+$x_s$,1/4,1/2-$z_s$ & 1/2+$x_s$,1/4,1/2	&	1/2+$x_s$,1/4,1/4	&-5/8,-3/4,1/4	&-3/8,-3/4,1/4
			\\
			\hline
			
		\end{tabular}	
	\end{table*} 

\appendix
\section{Determination of Immm as Photoinduced Lattice Instability}
\subsection{Supergroups of Pnma SnSe}
\begin{figure}
	\centering
	\includegraphics{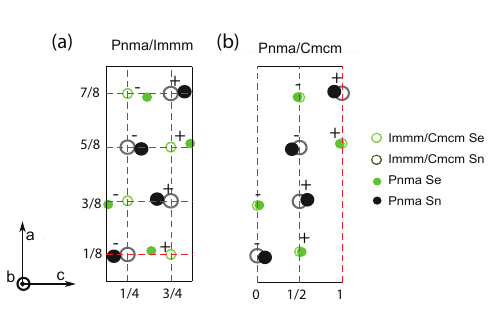}
	\caption{Comparison of the conventional Pnma unit cell with the Pnma/Immm and Pnma/Cmcm structures. Atoms of Pnma phase are marked with filled circles. Atoms in the Immm and Cmcm phase (open circle) are located in different high symmetry mirror planes (represented with dashed red lines). In all the structures, '+' means $y_{\text{s}}=\frac{3}{4}$, '-' means $y_{\text{s}}=\frac{1}{4}$. }  
	\label{fig:symmetry unit cell}
\end{figure}
The fractional positions of the Se and Sn atoms in the unit cell in the Pnma or higher symmetry structures can be specified with four free parameters $x_{\mathrm{Sn}}, z_{\mathrm{Sn}}, x_{\mathrm{Se}}, z_{\mathrm{Se}}$ with atoms located at $\pm(x_s,\frac{1}{4},z_s)$ and $\pm(\bar{x}_s+\frac{1}{2},\frac{3}{4},z_s+\frac{1}{2})$($s\in\{ \text{Sn}, \text{Se} \}$) corresponding to the Pnma Wyckoff site 4c\citep{Wyckoff}.

To find the relevant higher symmetry structures, we sort through all compatible structures as follows. As mentioned in the main text the early time dynamics can be considered effectively at constant volume, thus we search for orthorhombic structures that are supergroups (higher symmetry) of Pnma, whose atomic sites can be described with the Wyckoff 4c site of Pnma for certain values of $x_s$ and $z_s$ after a proper coordinate transformation. We also require that the direct linear displacement towards the candidate structure must match $\Delta x_{\rm Se}>0$, $\Delta z_{\rm Se}<0$, $\Delta x_{\rm Sn}>0$, $\Delta z_{\rm Sn}>0$ as observed in the experiment. Furthermore, we require that bond lengths $d_1 ,d_2 ,d_3$ do not change by more than 50\% of the original length. With these criteria, we search through the 230 space groups as well as all the possible Wyckoff sites in each space group\cite{Wyckoff,generalBilbao}, and find two supergroups that satisfy these constraints: Pmmn (Se:2d, Sn:2b) and Immm (Se:2d, Sn:2b).

\begin{figure}
	\centering
	\includegraphics{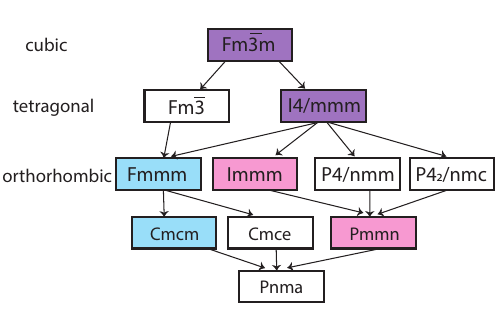}
	\caption{ Subgroup descendants of Fm$\bar{3}$m~\citep{subgroupsupergroup}. The descendants of Fm$\bar{3}$m are branched into the Fmmm (blue) and Immm (pink), which separately branches down to Cmcm and Pmmn, which have a common child of Pnma. Cmcm is not a subgroup of Immm. }  
	\label{fig:space group relation}
\end{figure}
\begin{figure}
	\centering
	\includegraphics{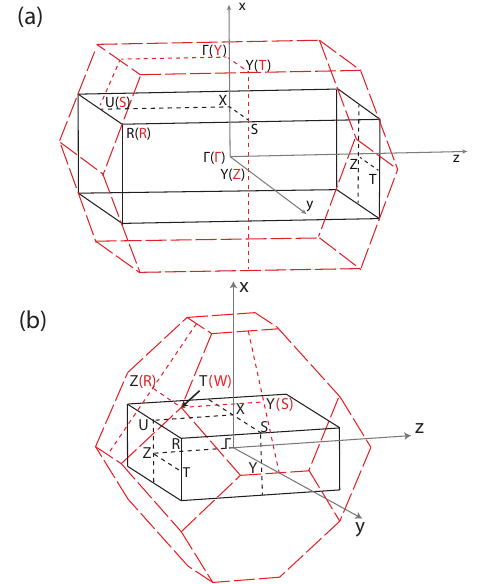}
	\caption{ The Brillouin zones of Pnma (black) and in red, Cmcm (a) and Immm (b). The red symbols denote the high symmetry points of the Brillouin zone of Cmcm (a) or Immm (b) and black symbols are those for Pnma. The $\mathbf{x,y,z}$ axes overlap with Pnma $\mathbf{a,b,c}$ axes respectively.}

	\label{fig:BZ}
\end{figure}

Pmmn is an intermediate structure between Pnma and Immm (see Fig.~\ref{fig:space group relation} for a group-subgroup relationship between the relevant structures). A symmetry enhancement from Pnma to Pmmn moves $z_s$ to the high symmetry values $z_{\rm Sn}=\frac{1}{4}$, $z_{\rm Se}=\frac{1}{4}$. A further symmetry enhancement from Pmmn towards Immm requires in addition that $x_{\rm Sn}=\frac{1}{8}$, $x_{\rm Se}=\frac{7}{8}$ (see Table ~\ref{tab:Wyckoff Site table}). 

The general structural transformation from space group Immm to Pnma involves two irreducible representations, LD$_4$ (at wave vector (0,0,$\frac{1}{2}$) in the Immm Brillouin zone, see Fig.~\ref{fig:BZ}), which distorts Pnma into the Pmmn structure, and X$_2^-$ (at wave vector (1,1,1) in the Immm Brillouin zone), which further distorts the Pmmn structure into Immm\citep{CORRparameter,representations}. In SnSe with its specific atomic sites, LD$_4$ describes the shear between bilayers along the $\bf{c}$-axis so that Sn and Se atoms are aligned along the $\bf{a}$-axis, while X$_2^-$ involves the motion along $\bf{a}$-axis so that Sn and Se atoms from the same atomic layer are aligned along the $\bf{c}$-axis, as can be seen from Fig.~\ref{fig:symmetry unit cell}. As pointed out in the main text, the A$_g^{(1)}$ mode of Pnma has a strong component of LD$_4$, and A$_g^{(2)}$ is close to X$_2^-$. Importantly, these two modes are the dominant components in the photoexcited atomic motion as shown in Fig.\ref{fig:reconstruction} (a). The DFT calculations presented in the main text (see  Fig.~\ref{fig:Sn_displacement}) further confirm the identification of Immm as the relevant symmetry. 

The coordinates listed in Table \ref{tab:Wyckoff Site table} are obtained from their respective standard Wyckoff positions by converting to a Pnma basis using the following transformation $(x,y,z)^T|_{\rm Pnma} = P [(x,y,z)^T+v]|_{\rm Cmcm, Immm}$ where for Immm (conventional unit cell)
\[
  P=
  \left[ {\begin{array}{ccc}
   0 & 0 & \frac{1}{2}\\
   0 & 1 & 0 \\
   -1 & 0 & 0 \\
  \end{array} } \right],
\] and $v=(\frac{1}{4} \frac{1}{4} \frac{1}{4})^T$.
For a Cmcm (conventional unit cell) to Pnma transformation, \[
  P=
  \left[ {\begin{array}{ccc}
   0 & 1 & 0\\
   0 & 0 & 1 \\
   1 & 0 & 0 \\
  \end{array} } \right],
\] and $v=(0 0 0)^T$~\citep{generalBilbao}. 

The space groups discussed in this paper (Pnma, Cmcm, Pmmn, Immm) are subgroup descendants of Fm$\bar{3}$m. Their relations are summarized in Fig~\ref{fig:space group relation}. Importantly, Cmcm and Immm do not have a supergroup-subgroup relation, thus one structure cannot be distorted into the other by a displacement that respects all of its symmetries (A$_{g}$ modes). All the discussed structures are parametrized with Pnma 4c Wyckoff sites ($x_{\mathrm{Sn}}, z_{\mathrm{Sn}}, x_{\mathrm{Se}}, z_{\mathrm{Se}}$) in Table~\ref{tab:Wyckoff Site table} and Fig.~\ref{fig:symmetry unit cell}. Structures with atoms occupying high symmetry positions can eliminate some of the four free parameters and lead to a reduction of the number of A$_{g}$ modes. Pnma has four A$_{g}$ modes. In the Cmcm structure two of the four A$_g$ modes become zone boundary modes and the unit cell is halved. In Immm all of the A$_g$ modes are eliminated and the unit cell is quartered. 
Throughout this paper, the real space fractional coordinates are referenced to the conventional Pnma unit cell, and the reciprocal space wavevectors are expressed in the Brillouin zone of the Pnma structure unless otherwise stated. 
\begin{figure*}
		\centering
		\includegraphics{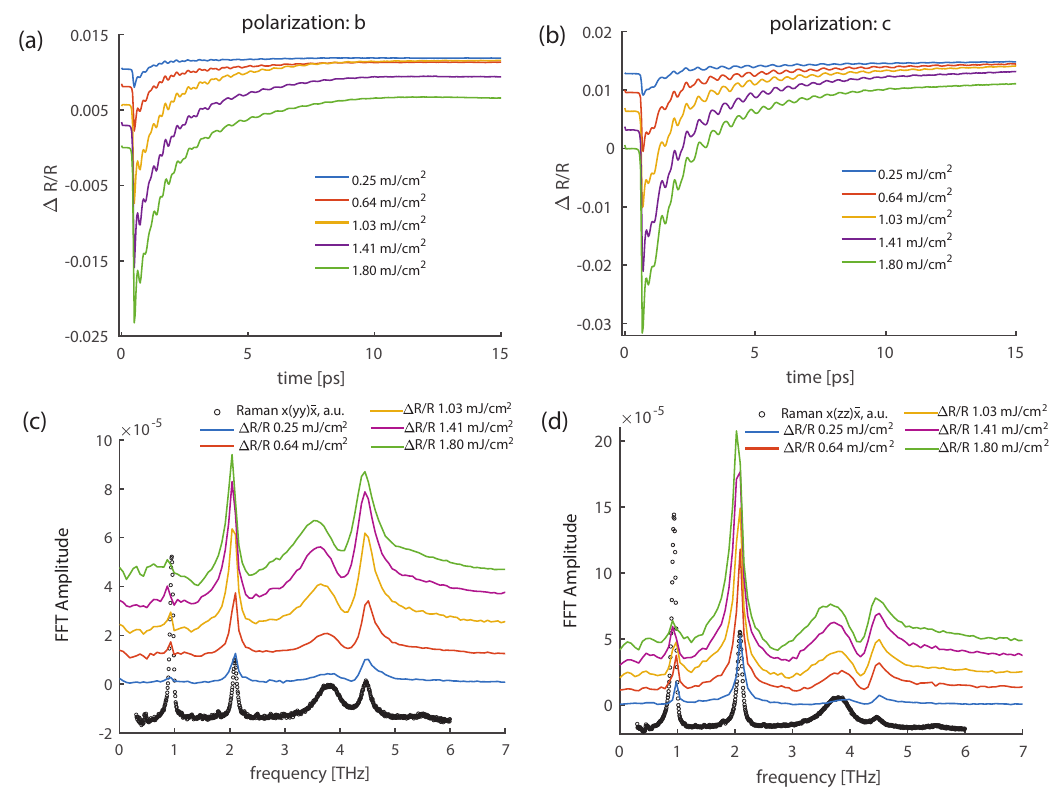}
		\caption{Raman spectrum and optical reflectivity. (a), (b) Optical reflectivity data, with both pump and probe beam polarized along $\bf{b}$, $\bf{c}$ respectively. The fluence range selected for reflectivity measurements matches that of the time resolved x-ray diffraction experiment. (c), (d) Raman spectrum under $x(yy)\bar{x}$, $x(zz)\bar{x}$ geometries (black circles), shown together with Fourier transform of pump probe reflectivity data in (a), (b) (colored lines). All four A$_g$ modes of SnSe are identified both in Raman measurement and pump probe optical reflectivity. }
		\label{fig:raman_optical}
	\end{figure*}

\begin{table*}
\caption{\label{tab:bond length} Bond length(${\AA}$) and bond angle($^{\circ}$) change between structures with fixed lattice constants ($a=11.31$\,\AA\, $b= 4.12$\,\AA\, $c=4.30$\,\AA\, of room temperature SnSe). 
Pnma equilibrium structure (Pnma$_{e.q.}$) takes the fractional coordinates as calculated by DFT. Cmcm has experimentally measured fractional coordinates at 855~K~\citep{sist2016crystal}. Immm fractional coordinates are as listed in Table~\ref{tab:Wyckoff Site table}. Bond length and angle changes for photoexcited SnSe are calculated based on Fig.~\ref{fig:reconstruction}~(b), with error bars propagated from the linear regression error of $b_i$ in the fit of $\alpha_i=b_i (\nu_2-\nu_2^0)$ to data in Fig.~\ref{fig:reconstruction} (a). 
See Fig.~\ref{fig:bond angles} for definitions of $\theta_{1-4}$.
The experimentally measured value with the same sign of change as $\Delta d_{\rm Immm}$ ($\Delta \theta_{\rm Immm}$) but opposite sign of change as $\Delta d_{\rm Cmcm}$ ($\Delta \theta_{\rm Cmcm}$) are highlighted in bold.}

\begin{ruledtabular}
\begin{tabular}{c|cc|cc|cc|c|cc|cc|cc}
& Pnma$_{e.q.}$&  \MyHead{1.8cm}{$\Delta d_{p.e.}\times100$} \vline& Immm & \MyHead{1.3cm}{$\Delta d_{\rm Immm}$}\vline & Cmcm & \MyHead{1.3cm}{$\Delta d_{\rm Cmcm}$} \vline   &  & Pnma & \MyHead{1.3cm}{$\Delta \theta_{p.e.}$} \vline& Immm &  \MyHead{1.3cm}{$\Delta \theta_{\rm Immm}$} \vline& Cmcm &  \MyHead{1.3cm}{$\Delta \theta_{\rm Cmcm}$} \\
\hline
$d_1$ &2.74  &\bf{1.00}(0.04)    &2.83 &\bf{0.09} & 2.63 & \bf{-0.11} & $\theta_{1}$ & 144.62 & \bf{0.33}(0.01) & 180 & \bf{35.38} & 144.62 & \bf{-0.00}\\
$d_2$ &2.79  &0.10(0.01) &2.98 &0.18 &  2.99 &0.19& $\theta_{2}$ & 156.54 & \bf{0.12}(0.01) & 180 & \bf{23.46}& 144.52 & \bf{-11.92}\\
$d_3$ &3.20  &-0.44(0.02) &2.98&-0.22&2.99&-0.22& $\theta_{3}$ & 80.37 &0.37(0.02)& 90 & 9.63& 85.83 &5.45\\
$d_4$ &3.36  &\bf{-1.94}(0.05) &2.83& \bf{-0.53}&3.71&\bf{0.35}& $\theta_{4}$ & 99.82 & -0.40(0.02)& 90 & -9.82 & 94.18 &-5.64\\

\end{tabular}
\end{ruledtabular}
\end{table*}
\begin{figure}
	\centering
	\includegraphics{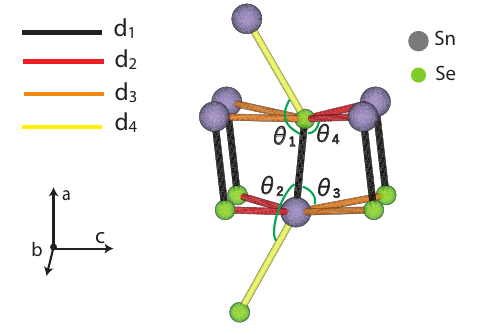}
	\caption{ Pnma SnSe bonds and bond angles. Atoms in the shown local structure of Pnma SnSe relative to the unit cell can be referenced from Fig.~\ref{fig:symmetry unit cell_maintext}. }
	\label{fig:bond angles}
\end{figure}

\subsection{Quantitative Analysis of Bond Angle and Bond Length Changes}
In Table~\ref{tab:bond length} we compare the experimental bond length and bond angle changes of the photoexcited structure with the values of the Pnma-Cmcm and Pnma-Immm structural distortions. The photoexcited $\Delta \theta$, $\Delta d$ quantities are calculated from displacements as shown in Fig.~\ref{fig:reconstruction} (b) without the magnifying factor $\times$100 (see Fig.~\ref{fig:bond angles} for definitions of bond angles). Changes towards the Immm and Cmcm phases are based on the fractional coordinates of the corresponding structures, albeit scaled by the Pnma lattice constant for a meaningful comparison with the photoexcited state with constrained lattice constants. 

The trend to remove the corrugation within a bilayer is mainly reflected in increased $\theta_3$ and decreased $\theta_4$, as well as the stretched $d_2$ and compressed $d_3$ (see Table.~\ref{tab:bond length}). In both Immm and Cmcm the intra-layer corrugation is reduced, explaining the same sign of $\Delta d_2$, $\Delta d_3$, $\Delta \theta_3$ and $\Delta \theta_4$ in the structural distortions towards Immm and Cmcm. Importantly, however, $\Delta d_1>0$, $\Delta d_4<0$, $\Delta \theta_1>0$ and $\Delta \theta_2<0$ are consistent with a distortion towards Immm but not with Cmcm, reflecting the fact that Immm restores the local quasi-octahedral symmetry while Cmcm does not. 
\section{Raman Spectroscopy and Pump-Probe Reflectivity Measurements}

Room temperature Raman measurements were performed with a continuous wave laser with photon energy of 1.96~eV (wavelength 633nm) with a Horiba LabRAM HR Evolution spectrometer. 
The spontaneous Raman spectrum was taken under $x(yy)\bar{x}$ and $x(zz)\bar{x}$ geometries where the incident beam and reflected beams have the same polarization. The room temperature pump-probe reflectivity measurements were performed using a Coherent RegA Ti:sapphire laser system with a repetition rate of 250~kHz with photon energy of 1.55~eV (800 nm) and pulse duration of 46~fs for both pump and probe. Pump-probe measurement was performed under normal incidence with incoming beam and reflected beam propagating along crystal $\bf{a}$-axis, and with both pump and probe polarized along either $\bf{b}$-axis or $\bf{c}$-axis. For both spontaneous Raman and pump-probe measurements, we used the same single crystal sample as that used in time-resolved x-ray diffraction experiment. 

Fig.\ref{fig:raman_optical} (a) and (b) show time-resolved optical reflectivity of SnSe with polarizations along $\bf{b}$ and $\bf{c}$. (c) and (d) show the Fourier transforms of (a) and (b), as well as the Raman spectrum under $x(yy)\bar{x}$ and $x(zz)\bar{x}$. Our Raman spectrum is consistent with previous measurements~\citep{chandrasekhar1977infrared}. 

The frequency of the A$_g^{(2)}$ mode measured in Raman spectroscopy is 2.08~THz. If we allow the $\nu_2^0$ to be a fitting parameter in the fit $\alpha_i=b_i (\nu_2-\nu_2^0)$ to data in Fig.~\ref{fig:reconstruction} (a), $\nu_2^0\approx2.07$~THz.
The $\text{A}_g^{(3)}$ mode around 3.8 THz is clear in these optical measurements, while it was not detected above the noise in time-resolved x-ray diffraction data. The drastic softening of A$_g^{(3)}$, a mode that is mainly polarized in the $\mathbf{b-c}$ plane (Fig.~\ref{fig:raw data}), can be related to the strengthening of the resonant bonds that also result in the large softening of A$_g^{(3)}$ as observed in the diffraction results. See Appendix C for discussion of photoexcited forces.

\begin{figure*}
	\centering
	\includegraphics{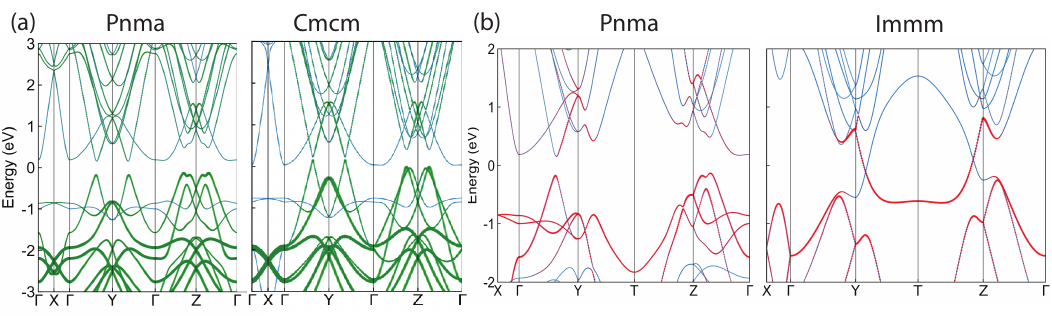}
	\caption{(a) Equilibrium Pnma phase and Cmcm phase electron band structure. Thickness of green lines represent projection to Se $p_{y,z}$ orbitals. (b) Equilibrium Pnma phase ($N_{h}$ = 0.0\,hole/f.u.) and photoexcited Immm structure ($N_{h}$=0.20\,hole/f.u.) electron band structure. Thickness of red lines represent projection to Sn 5$s$ orbitals.}
	\label{fig:Se p bands}
\end{figure*}

\begin{figure}
	\centering
	\includegraphics{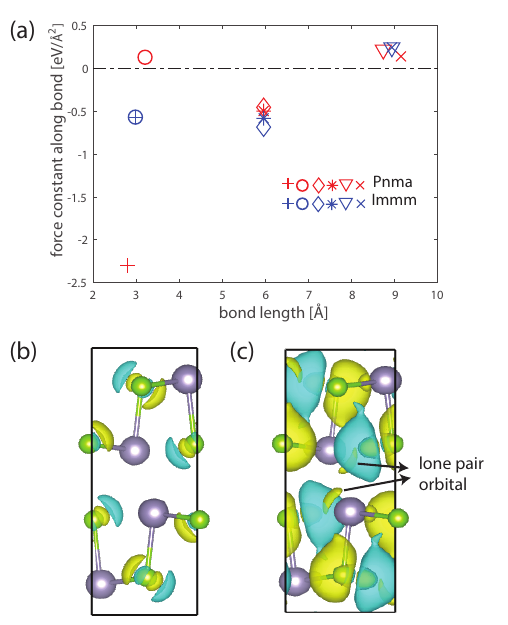}
	\caption{(a) Force constants along the bonding direction, shown for bonds of the resonant bonded network. Immm structure corresponds to hole doping DFT at $N_{h}$=0.20\,hole/f.u. Bonds are distinguished by different markers. (b-c) Photoexcited state differential charge isosurface (0.022 \,electron/\AA$^3$) plot, for $N_{h}$=0.175\,hole/f.u.(b) and for $N_{h} $=0.20\,hole/f.u.(c). Atoms in the unit cells are of the Pnma equilibrium structure. The yellow isosurface represents a differential (photoexcited subtracted by equilibrium) positive charge (negative EDOS), while blue isosurface represents a negative charge difference (positive EDOS).}
	\label{fig:resonant bonds}
\end{figure}
\begin{figure}
	\centering
	\includegraphics{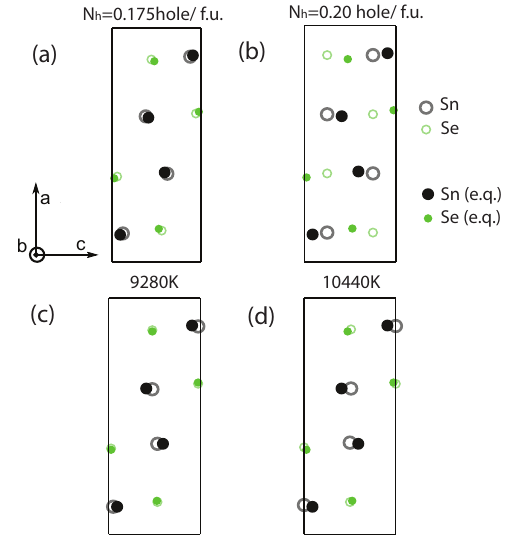}
	\caption{Comparison between hole doped DFT and DFT with modified electronic temperatures. The Sn (Se) atoms are empty grey (green) circles for structures with modified electronic structures, and filled circles for the equilibrium. (a,b) Crystal structures with the increasing number of holes per SnSe formula unit $N_{h}$. Atomic positions are displaced towards Immm as the excitation density increases. (c,d) DFT calculations by raising the electronic temperatures as indicated with the title for each plot, showing atomic displacements towards Cmcm as the temperatures increase.}
	\label{fig:simulation}
\end{figure}

\begin{table}
		\caption {\label{tab:force constants} Selected elements of the force constants $k_{n,ij}$ for the atom pairs connected by the $d_n$ ($n=1-4$) bonds in the Pnma, Immm and Cmcm structures. The Cmcm values are taken from~\citep{hong2019phase}, in which the $d_4$ values are not provided. }  
		\centering
		\begin{tabular}{c|c|c|c}
			\hline
			$k$ (eV$/\text{\AA}^2)$ &  Pnma & Immm &Cmcm
			\\
			\hline
		    $k_{1,xx}$ 	& -3.31	&	-1.66 & -3.92
			\\
			$(k_{2,yy}+k_{2,zz})/2$ 	& -1.32	&	-0.18	&-0.35
			\\
			$(k_{3,yy}+k_{3,zz})/2$ 	& 0.01&	-0.18	&-0.35
			\\
			$k_{4,xx}$ 	& 0.08	&	-1.66	& -
			\\
			$(k_{4,yy}+k_{4,zz})/2$ 	& -0.44	&	-0.08	& -
			\\
			\hline
			
		\end{tabular}	
\end{table}
\section{Density Functional Theory}
DFT (and hole doped DFT) was performed using VASP, with the projected-augmented-wave (PAW) and local density approximation (LDA)~\citep{kresse1996efficient, blochl1994projector, kresse1999ultrasoft}, which proves to yield accurate phonon dispersions~\citep{li2015orbitally, bansal2016phonon, Lanigan2020} and provides better agreement with INS and Raman measurements than the Perdew-Burke-Ernzerhof (PBE) generalized gradient approximation (GGA)~\citep{bansal2016phonon, Lanigan2020}. The Pnma equilibrium structure is relaxed with a kinetic energy cutoff of 500~eV and an electronic Monkhorst-Pack grid with $6 \times 12 \times 12$ $k$-points, giving lattice constants ($a=11.31$\,\AA\, $b= 4.12$\,\AA\, $c=4.30$\,\AA\,) in good agreement with x-ray diffraction experimental report at 296~K: \citep{wiedemeier1979thermal} ($a=11.50$\,\AA\, $b=4.16$\,\AA\, $c=4.45$\,\AA\,) and \citep{sist2016crystal} ($a=11.44$\,\AA\, $b=4.13$\,\AA\, $c=4.45$\,\AA\,). 
We use the DFT calculated fractional coordinates ($x_{\text{Se}}=0.14, z_{\text{Se}}=0.48, x_{\text{Sn}}=0.88, z_{\text{Sn}}=0.09$) for the equilibrum Pnma structure, which are also in good agreement with \citep{sist2016crystal} ($x_{\text{Se}}=0.12, z_{\text{Se}}=0.48, x_{\text{Sn}}=0.86, z_{\text{Sn}}=0.11$). 
In constrained-DFT, the structure optimization was performed on the $2 \times 4 \times 4$ supercell of Pnma conventional unit cell, using a $3 \times 3 \times 3$ $k$-point mesh, 500~eV energy cutoff. 
In both calculations methods, the lattice constants were fixed to those of the equilibrium Pnma structure.

Fig.~\ref{fig:Se p bands} (a) shows Se 4$p_{y,z}$ orbital projected bands in both Pnma and Cmcm phases. The band gap is closed along $\Gamma-Y$ by bands mainly composing of Se 4$p_{y,z}$ (hybridized with Sn 5$s$) in a phase transition from Pnma to Cmcm, which reveals the Peierls nature of the lattice instability~\citep{li2015orbitally,hong2019phase}. The non-bonding lone-pair orbital is a hybridized orbital of Sn 5$s$ and Se 4$p_x$, the electron bands projection onto the latter are shown in the main text. For reference, we also plot the Sn 5$s$ orbital projected Pnma and Immm band structure in Fig.~\ref{fig:Se p bands} (b).

The calculated force constants of relevant bonds are listed in Table~\ref{tab:force constants}. The inter-layer $d_4$ bond has the strength of within an order of magnitude of intra-layer bonds $d_1$ and $d_2$, reflecting the fact that SnSe is not a strongly 2D material~\citep{wang2018defects}. The $\mathbf{a}$ component of the $d_4$ force constant turns from weakly repulsive (positive) in Pnma to attractive (negative) in Immm. 
In Fig.~\ref{fig:resonant bonds} we show the force constants of the six selected nearest bonds along the $\bf{b-c}$ plane $p$-orbital network.
The forces for Immm were computed for $N_{h} $=0.20\,hole/f.u.
Below the excitation threshold $N_{h}$=0.20~hole/f.u., the electron density isosurface plot in Fig.~\ref{fig:resonant bonds} (b) ($N_{h}$=0.175~hole/f.u.) features hole doping into the $\bf{b-c}$ plane $p$ orbitals that are closer to the Fermi level than the lone pair orbitals. The electron density change of lone pair orbital (around the Sn atom in Fig.~\ref{fig:resonant bonds} (c)) is manifest at a higher excitation density $N_{h}$=0.20~hole/f.u., which has a smaller isosurface than the $\bf{b-c}$ plane $p$ orbitals but a larger effect in determining the structural distortion. 
\section{Linear Prediction and Phonond Mode Decomposition}
Linear prediction decomposes the data into a sum of exponentially decaying harmonic oscillators and decaying exponentials~\citep{barkhuijsen1985retrieval}. It has been applied to the analysis of time domain data, such as NMR spectra~\citep{barkhuijsen1985retrieval,led1991application}. Unlike nonlinear least squares, this method does not require the initial guess of fitting parameters. It is a convex optimization problem, and thus, will not be stuck in a local minimum as in a least square fit. Furthermore, linear prediction can give a statistically sound estimate of the number of oscillators contained in the signal~\citep{epps2019singular1,epps2019singular2}, which is usually an external input in a least square fit. 
\begin{figure*}
	\centering
	\includegraphics{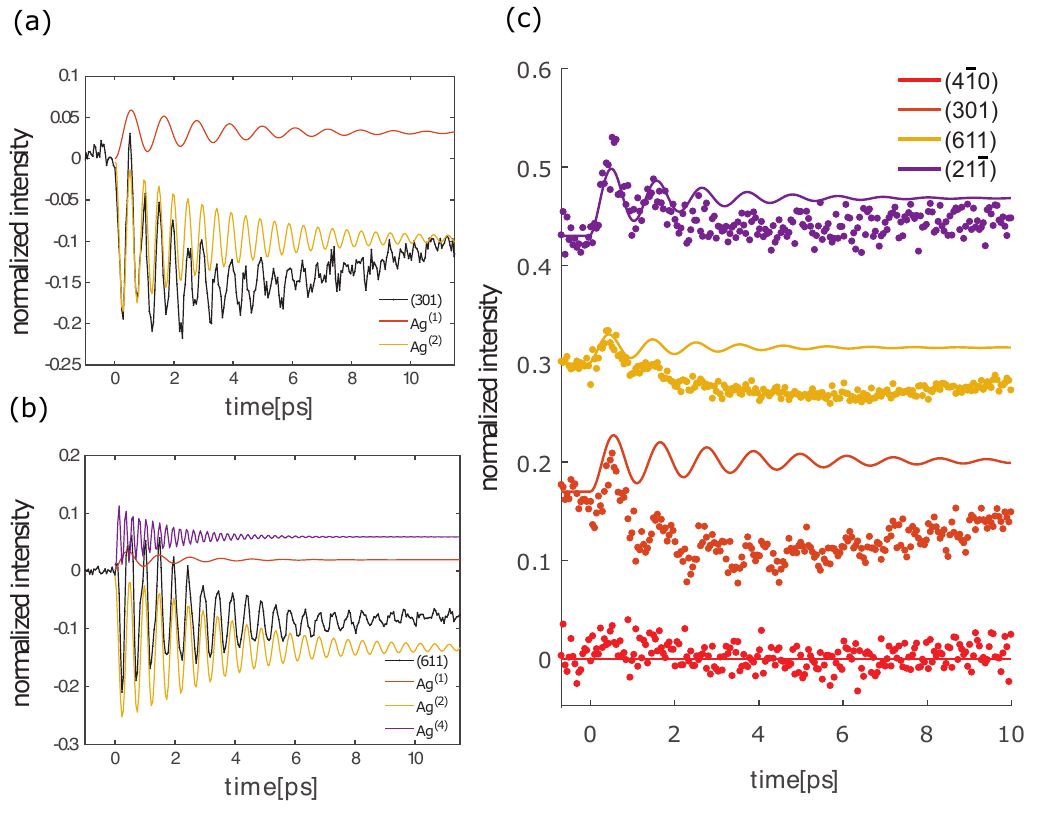}
	\caption{\label{fig:appendix_mode_decomp} { (a) Linear prediction showing decomposition of modes for peak (611), and similarly (b) for peak (301). The black traces are the raw data, and the colored traces are the components of A$_g$ modes. (c)  Color dots represent the residual of raw data  after subtracting off the A$_g^{(2)}$ and A$_g^{(4)}$ components, the solid color lines represent the A$_g^{(1)}$ component.}  }
	
\end{figure*}
Similar to Fig.\ref{fig:Ag1 mode} (a) for (21$\bar{1}$) Bragg peak, Fig.\ref{fig:appendix_mode_decomp} (a) and (b) show the decomposition prescribed by Eq.(\ref{eq:intensity of DECP decomposition}) for the (611) and (301) Bragg peaks. The black trace is the experimental data and the colored lines are the DECP components. In (301) the A$_g^{(4)}$ component is not visible. 
Fig.\ref{fig:appendix_mode_decomp} (c) shows an analysis of the isolated A$_g^{(1)}$ component for all four peaks (21$\bar{1}$), (611), (301) and (4$\bar{1}$0). Here we show the residual (colored dots) of the experimental intensity subtracted by the components of A$_g^{(2)}$ and A$_g^{(4)}$, as well as the linear predicted A$_g^{(1)}$ component (colored lines). The residual is of course very noisy since modes with high signal level are subtracted. Nonetheless the initial phase of the A$_g^{(1)}$ are manifest in both the residual and the linear predicted DECP component. We note that peak (4$\bar{1}$0) is not sensitive to mode A$_g^{(1)}$. 
\bibliography{SnSe}

\end{document}